# Spin and orbital angular momenta of acoustic beams

Konstantin Y. Bliokh[1,2] and Franco Nori[1,3]

[1]*Theoretical Quantum Physics Laboratory, RIKEN Cluster for Pioneering Research, Wako-shi, Saitama 351-0198, Japan*
[2]*Nonlinear Physics Centre, RSPE, The Australian National University, ACT 0200, Australia*
[3]*Physics Department, University of Michigan, Ann Arbor, Michigan 48109-1040, USA*

We analyze spin and orbital angular momenta in monochromatic acoustic wave fields in a homogeneous medium. Despite being purely longitudinal (curl-free), inhomogeneous acoustic waves generically possess nonzero spin angular momentum density caused by the local rotation of the vector velocity field. We show that the integral spin of a localized acoustic wave vanishes in agreement with the spin-0 nature of longitudinal phonons. We also show that the helicity or chirality density vanishes identically in acoustic fields. As an example, we consider nonparaxial acoustic Bessel beams carrying well-defined integer orbital angular momentum, as well as nonzero local spin density, with both transverse and longitudinal components. We describe the nontrivial polarization structure in acoustic Bessel beams and indicate a number of observable phenomena, such as nonzero energy density and purely-circular transverse polarization in the center of the first-order vortex beams.

## 1. Introduction

Spin and orbital angular momenta are important properties of both classical light and photons, which have been extensively studied in the past few decades [1–5]. It is well-known that the spin angular momentum (AM) of photons originates from the circular (or elliptical) polarization of electromagnetic waves, i.e., rotation of the electric or magnetic fields, while the orbital AM is produced by the circulation of the phase gradient, i.e., canonical momentum density in optical fields. Optical beams or photon states carrying longitudinal (i.e., along the propagation direction) spin and orbital angular momenta have become ubiquitous in many areas of modern optics, including optical manipulations, nano-photonics, quantum information, etc. Recently, it was noticed that in inhomogeneous optical fields (e.g., nonparaxial or evanescent), rotations of the electric and magnetic fields also occur in the propagation plane, which results in a *transverse spin* [5–13] with unusual properties, such as spin-momentum locking via evanescent waves [14–20].

Acoustic and elastic waves or phonons exhibit many phenomena similar to those in optics and electromagnetism. It has been known that polarization/spin properties of transverse (shear, divergence-free) elastic waves [21] are quite similar to those of light, and transverse phonons can be regarded as spin-1 particles analogous to photons [22–27]. Furthermore, since orbital AM is a phase property largely independent of the polarization (at least in paraxial fields), it has been intensively studied, both theoretically and experimentally, in acoustic vortex beams similar to the optical ones [28–37]. These acoustic beams in fluids or gases were essentially treated as *scalar* pressure waves. In fact, in terms of the *vector velocity field*, these are *longitudinal* (i.e., curl-free) *vector* waves corresponding to spin-0 longitudinal phonons [21].

Finally, very recently, it was noticed that longitudinal vector fields, including acoustic waves, can also have nontrivial polarization and spin properties [38–40]. Even though acoustic plane waves are always linearly polarized (velocity is aligned with the wavevector), inhomogeneous waves (e.g., evanescent or interference fields) locally exhibit *elliptical* polarization with rotating velocity vector. This generates non-zero *spin AM density* in curl-free acoustic fields. It was shown that the transverse spin in inhomogeneous acoustic waves [38–40] is quite similar to that in optical fields [5–20], and this phenomenon was observed experimentally [39].

In this work, continuing the studies in [38–40], we provide general analysis of the spin and orbital angular momenta, as well as other dynamical quantities (energy flux, canonical momentum, and helicity) in monochromatic acoustic fields. We show that the *helicity (chirality) density vanishes identically* in longitudinal wave fields, while the spin AM density is generically non-zero. Furthermore, the *integral spin AM* of a localized wave field also *vanishes*, which agrees with the spin-0 nature of longitudinal phonons. Due to the presence of spin density, the energy flux density (i.e., the kinetic momentum density analogous to the Poynting vector) differs from the canonical momentum density. The latter determines the orbital AM of the acoustic field. We compare all dynamical characteristics of acoustic fields with their electromagnetic counterparts. As an explicit example of acoustic field carrying both spin and orbital angular momenta, we consider nonparaxial vortex Bessel beams. We show that such beams have rather nontrivial polarization properties, with spin AM having both transverse (azimuthal) and longitudinal components, purely-circular vortex-induced transverse polarization in the beam center, and non-zero energy density in the center of the first-order vortex beams.

## 2. General properties of acoustic wave fields

We start with the linear equations for acoustic (sound) waves in a homogeneous dense medium, fluid or gas [41]:

$$\beta \frac{\partial P}{\partial t} = -\nabla \cdot \mathbf{v}, \qquad \rho \frac{\partial \mathbf{v}}{\partial t} = -\nabla P. \tag{1}$$

Here the variables are: the velocity $\mathbf{v}(\mathbf{r},t)$ and the pressure $P(\mathbf{r},t)$ fields, while the real-valued medium parameters are: the mass density $\rho$ and the compressibility $\beta = 1/B$ ($B$ is the bulk modulus). Equations (1) obey the energy conservation law, an acoustic analogue of the electromagnetic Poynting theorem:

$$\frac{\partial}{\partial t}\left(\frac{\beta P^2}{2} + \frac{\rho \mathbf{v}^2}{2}\right) + \nabla \cdot (P\mathbf{v}) = 0, \tag{2}$$

where the expressions in the first and second parentheses determine the acoustic energy density and energy flux density, respectively [41].

From now on, we consider monochromatic acoustic waves of frequency $\omega$. Making the substitution $\mathbf{v}(\mathbf{r},t) \to \mathrm{Re}\left[\mathbf{v}(\mathbf{r})e^{-i\omega t}\right]$ and $P(\mathbf{r},t) \to \mathrm{Re}\left[P(\mathbf{r})e^{-i\omega t}\right]$, Eqs. (1) are reduced to the following equations for the *complex* velocity and pressure fields $\mathbf{v}(\mathbf{r})$ and $P(\mathbf{r})$:

$$\nabla \cdot \mathbf{v} = i\beta\omega P, \qquad \nabla P = i\rho\omega \mathbf{v}. \tag{3}$$

Equations (1) or (3) support only longitudinal (i.e., curl-free) waves: $\nabla \times \mathbf{v} = 0$. For plane waves with the wavevector $\mathbf{k}$, $\nabla \to i\mathbf{k}$, the dispersion relation and the "longitudinality" condition follow from Eqs. (3):

$$\omega^2 = k^2 c^2 \equiv \frac{k^2}{\rho\beta}, \qquad \mathbf{k} \times \mathbf{v} = 0, \tag{4}$$

where $c$ is the speed of sound.

Importantly, although commonly classified as "scalar waves", sound waves also have inherent *vector* properties [38–40]. Indeed, these waves are described by one scalar (pressure) and one vector (velocity) fields, which determine the qualitatively-different degrees of freedom in the acoustic field. These scalar and vector degrees of freedom are equally important, as can be seen from their equal contributions to the energy conservation law (2). In quantum-like terms, one can say that acoustic waves are described by the four-component "wavefunction" $\psi = (P, \mathbf{v})^T$. In what follows, we will use a fruitful analogy with electromagnetic waves described by Maxwell equations. The main difference is that Maxwell waves are described by two vector fields (electric and magnetic), $\psi = (\mathbf{E}, \mathbf{H})^T$, and these are transverse (i.e., divergence-free) rather than longitudinal: $\nabla \cdot \mathbf{E} = \nabla \cdot \mathbf{H} = 0$.

Similarly to electromagnetic waves [5,6,42,43], the main dynamical properties of acoustic wave fields are: the *energy*, *momentum*, and *angular momentum*. The time-averaged energy density and energy flux density (an acoustic counterpart of the Poynting vector) in a monochromatic acoustic field follow from Eq. (2):

$$W = \frac{1}{4}\left(\beta |P|^2 + \rho |\mathbf{v}|^2\right), \quad \mathbf{\Pi} = \frac{1}{2}\mathrm{Re}\left(P^* \mathbf{v}\right). \tag{5}$$

Employing the quantum-like formalism [5,42,43], the energy density can be regarded as the local expectation value of the energy (frequency) operator $\omega$, $W = \left(\psi | \omega | \psi\right)$, where the inner product $\left(\psi | \psi\right)$ is defined with the scaling coefficients $\beta / 4\omega$ and $\rho / 4\omega$ at the pressure and velocity degrees of freedom, respectively. Using this formalism, similarly to the electromagnetic case [5,6,8,10,42–44], we introduce the *canonical momentum density* of the acoustic field as the local expectation value of the momentum operator $\hat{\mathbf{p}} = -i\nabla$:

$$\mathbf{p} = \frac{1}{4\omega}\mathrm{Im}\left[\beta P^* \nabla P + \rho \mathbf{v}^* \cdot (\nabla)\mathbf{v}\right], \tag{6}$$

where $\left[\mathbf{v}^* \cdot (\nabla)\mathbf{v}\right]_i \equiv \sum_j v_j^* \nabla_i v_j$. The momentum density (6) represents the natural definition of the *local phase gradient* (i.e., the local wavevector) in a multicomponent field $\psi$ (for a single-component scalar field it would be proportional to $\nabla \mathrm{Arg}(\psi)$) [44].

In analogy with electromagnetism, the energy flux density (5) can also be associated with the momentum density (multiplied by $c^2$), but this should be regarded as the *kinetic momentum density* $\mathbf{\Pi}/c^2$ [41]. Using some vector algebra involving the "longitudinality" condition $\nabla \times \mathbf{v} = 0$, the difference between the kinetic and canonical momentum can be written as

$$\frac{\mathbf{\Pi}}{c^2} = \mathbf{p} + \frac{1}{4}\nabla \times \mathbf{S}, \quad \mathbf{S} = \frac{\rho}{2\omega}\mathrm{Im}\left(\mathbf{v}^* \times \mathbf{v}\right). \tag{7}$$

Here, $\mathbf{S}$ is the *spin AM density* of the acoustic waves [39,40]. Thus, entirely similar to the electromagnetic case [5,6,8,10,42,43], the difference between the kinetic and canonical momentum densities in the acoustic field is related to the presence of the spin AM density. This difference can be regarded as the *spin momentum* density $\mathbf{p}_S = \frac{1}{4}\nabla \times \mathbf{S}$ [8,10,42,44,45]. The only distinction as compared to electromagnetism is the pre-factor $1/4$ instead of $1/2$; this is because the scalar (pressure) part of the "wavefunction" $\psi$ does not contribute to the difference between the kinetic and canonical momentum. Remarkably, the spin density (7) can also be presented as the local expectation value of the spin-1 operator $\hat{\mathbf{S}}$ acting on the vector (velocity) degrees of freedom such that $\mathbf{v}^* \cdot (\hat{\mathbf{S}}) \mathbf{v} = \text{Im}(\mathbf{v}^* \times \mathbf{v})$ [5,42–44]: $\mathbf{S} = 2\langle \psi | \hat{\mathbf{S}} | \psi \rangle$, where the factor of 2 originates from the same asymmetry between the scalar and vector degrees of freedom [40].

We now describe the angular momentum of the acoustic field. The spin density (7) represents its intrinsic (i.e., origin-independent) part, while the extrinsic *orbital AM density* is determined by the canonical momentum (6) [5,6,8,10,42,43]:

$$\mathbf{L} = \mathbf{r} \times \mathbf{p}, \quad \mathbf{J} = \mathbf{L} + \mathbf{S}, \quad (8)$$

where $\mathbf{J}$ is the total AM density in the canonical picture. In the kinetic picture, the total AM density is extrinsic and is determined by the kinetic momentum: $\mathbf{M} = \mathbf{r} \times \mathbf{\Pi}/c^2$ [29,30,32]. Akin to the electromagnetic case, the equivalence between the canonical and kinetic pictures can be seen when considering *integral* over the whole space values of the above dynamical quantities, $\langle ... \rangle \equiv \int ... dV$, for a localized (i.e., sufficiently fast decaying at infinity) acoustic field. Namely, the volume integral of any localized solenoidal field vanishes, and

$$\left\langle \frac{\mathbf{\Pi}}{c^2} \right\rangle \equiv \langle \mathbf{p} \rangle, \quad \langle \mathbf{M} \rangle = \langle \mathbf{J} \rangle = \langle \mathbf{L} \rangle, \quad \langle \mathbf{S} \rangle = 0. \quad (9)$$

The last equality here follows from the potential character of the velocity field, $\mathbf{v} = -i(\rho\omega)^{-1} \nabla P$, and Eq. (7):

$$\mathbf{S} = \frac{1}{2\rho\omega^3} \text{Im}(\nabla P^* \times \nabla P) = \frac{1}{2\rho\omega^3} \nabla \times \text{Im}(P^* \nabla P). \quad (10)$$

Thus, the acoustic spin AM density is a *solenoidal* field, which can be regarded as the *vorticity* of the scalar part of the momentum density $\propto \text{Im}(P^* \nabla P)$ [44], and its integral value (9) *vanishes*. This corresponds to the well-known fact that longitudinal phonons are spin-0 particles. Nonetheless, the local spin AM density is generally nonzero in inhomogeneous acoustic fields and is an observable quantity [39,40].

Finally, we consider one more quantity, which plays an important role in Maxwell electromagnetism. This is the *helicity*, which is described by the projection of the spin onto the momentum [5,42,46–49]. Using the spin operator $\hat{\mathbf{S}}$ and the canonical momentum operator $\hat{\mathbf{p}}$, the helicity operator becomes proportional to the curl operator: $\hat{\mathfrak{S}} \propto \hat{\mathbf{S}} \cdot \hat{\mathbf{p}} = \nabla \times$ [48,49]. Due to the longitudinal character of acoustic waves, the helicity density *vanishes identically*:

$$\mathfrak{S} \propto \mathbf{v}^* \cdot \hat{\mathfrak{S}} \mathbf{v} \propto \mathbf{v}^* \cdot (\nabla \times \mathbf{v}) \equiv 0. \quad (11)$$

Since the helicity can be regarded as a measure of the *chirality* of the field, this means that acoustic waves *cannot* be chiral (even locally), and the absorption of longitudinal phonons cannot distinguish between chiral enantiomers interacting with acoustic fields (cf. the case of photons [49–52]).

Equations (6)–(11) are the main general results of our work. Their comparison with the analogous well-studied equations for electromagnetic waves in an isotropic medium with permittivity $\varepsilon$ and permeability $\mu$ is shown in the Table I.

Note that we used the spin-orbital canonical decomposition of the momentum and AM based on the canonical momentum density (6), which is *"democratic" (i.e., symmetric) with respect to the pressure and velocity degrees of freedom*. This is similar to the symmetric form of the energy and energy-flux densities (5) and to the dual-symmetric (with respect to the electric and magnetic contributions) approach to Maxwell electromagnetism [5,6,42–44,47,52]. Alternatively, one can introduce asymmetric "pressure-biased" and "velocity-biased" decompositions, which can be convenient for some problems, akin to the electric- and magnetic-biased quantities in electromagnetism [42,44,45]. Introducing the corresponding biased canonical momentum densities, $\mathbf{p}^{(P)} = \frac{\beta}{2\omega} \text{Im}[P^* \nabla P]$ and $\mathbf{p}^{(v)} = \frac{\rho}{2\omega} \text{Im}[\mathbf{v}^* \cdot (\nabla)\mathbf{v}]$, so that $\mathbf{p} = \frac{1}{2}(\mathbf{p}^{(v)} + \mathbf{p}^{(P)})$, then the biased spin-orbital decompositions become:

$$\frac{\mathbf{\Pi}}{c^2} = \mathbf{p}^{(v)} + \frac{1}{2}\nabla \times \mathbf{S} = \mathbf{p}^{(P)}. \tag{12}$$

This shows that the spin momentum density becomes twice as large in the velocity-biased approach and vanishes in the pressure-biased approach: $\mathbf{p}_S^{(v)} = \frac{1}{2}\nabla \times \mathbf{S}$ and $\mathbf{p}_S^{(P)} = 0$. Nonetheless, the spin AM density (7) is still well-defined because it corresponds to the real mechanical AM produced by the local elliptical motion of the medium particles [38–40]. Different relations between the spin AM and spin momentum densities do not cause any contradiction because of the integral connection between these quantities and the spin-0 nature of the field: $\langle \mathbf{S} \rangle = \langle \mathbf{r} \times \mathbf{p}_S \rangle = \langle \mathbf{r} \times \mathbf{p}_S^{(P)} \rangle = \langle \mathbf{r} \times \mathbf{p}_S^{(v)} \rangle = 0$. In contrast, in electromagnetism, the relation between the spin momentum and AM densities is fixed as $\mathbf{p}_S = \frac{1}{2}\nabla \times \mathbf{S}$, to produce $\langle \mathbf{S} \rangle = \langle \mathbf{r} \times \mathbf{p}_S \rangle \neq 0$ [42].

| | **Acoustics** | **Electromagnetism** |
|---|---|---|
| Fields | velocity $\mathbf{v}$, pressure $P$ | electric $\mathbf{E}$, magnetic $\mathbf{H}$ |
| Constraints | $\nabla \times \mathbf{v} = 0$ | $\nabla \cdot \mathbf{E} = 0$, $\nabla \cdot \mathbf{H} = 0$ |
| Energy density | $\frac{1}{4}(\rho|\mathbf{v}|^2 + \beta|P|^2)$ | $\frac{1}{4}(\varepsilon|\mathbf{E}|^2 + \mu|\mathbf{H}|^2)$ |
| Canonical momentum density | $\frac{1}{4\omega}\text{Im}[\beta P^* \nabla P + \rho \mathbf{v}^* \cdot (\nabla)\mathbf{v}]$ | $\frac{1}{4\omega}\text{Im}[\varepsilon \mathbf{E}^* \cdot (\nabla)\mathbf{E} + \mu \mathbf{H}^* \cdot (\nabla)\mathbf{H}]$ |
| Kinetic momentum density | $\frac{1}{2c^2}\text{Re}(P^* \mathbf{v}) = \mathbf{p} + \frac{1}{4}\nabla \times \mathbf{S}$ | $\frac{1}{2c^2}\text{Re}(\mathbf{E}^* \times \mathbf{H}) = \mathbf{p} + \frac{1}{2}\nabla \times \mathbf{S}$ |
| Spin AM density | $\frac{1}{2\omega}\rho \text{Im}(\mathbf{v}^* \times \mathbf{v})$ | $\frac{1}{4\omega}[\varepsilon \text{Im}(\mathbf{E}^* \times \mathbf{E}) + \mu \text{Im}(\mathbf{H}^* \times \mathbf{H})]$ |
| Orbital AM density | $\mathbf{L} = \mathbf{r} \times \mathbf{p}$ | $\mathbf{L} = \mathbf{r} \times \mathbf{p}$ |
| Integral AM values | $\langle \mathbf{M} \rangle = \langle \mathbf{L} \rangle$, $\langle \mathbf{S} \rangle = 0$ | $\langle \mathbf{M} \rangle = \langle \mathbf{L} \rangle + \langle \mathbf{S} \rangle$, $\langle \mathbf{S} \rangle \neq 0$ |
| Helicity | $\mathfrak{S} \equiv 0$ | $\mathfrak{S} \neq 0$ |

**Table I.** Comparison of acoustic and electromagnetic quantities and properties.

## 3. Acoustic Bessel beams

We now apply the above general theory to the explicit example of acoustic Bessel beams. Bessel beams have been repeatedly considered in optics [53–58], quantum physics [59–62], and acoustics [32,63–65] as a convenient example of nonparaxial vortex beams carrying both spin and orbital angular momenta and allowing a laconic analytical description. It is known that the $z$-propagating Bessel beam of order $\ell = 0, \pm 1, \pm 2,...$ represents a superposition of plane waves with the same frequency and wavevectors $\mathbf{k}$ uniformly distributed over a circle with fixed polar angle $\theta_0$ (aperture angle), Fig. 1(a). These plane waves have mutual phases which depend on the azimuthal angle $\phi$ in $\mathbf{k}$-space as $\exp(i\ell\phi)$, producing the helical phase (vortex) and the orbital AM in the in real-space beam, Fig. 1(b) [1–5].

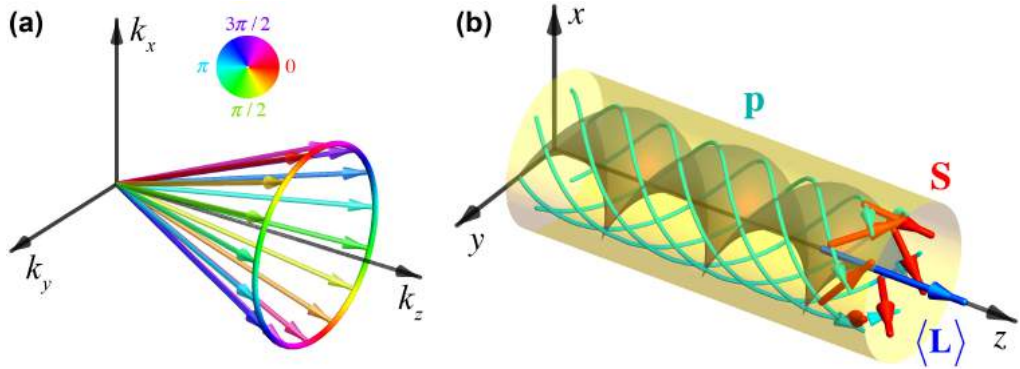

**Fig. 1.** Schematics of the acoustic Bessel beams. **(a)** The momentum (plane-wave) spectrum of the beam is a circle with fixed polar angle $\theta_0$. The mutual phases of the plane waves (color-marked) have an azimuthal gradient and the $2\pi\ell$ increment around the circle ($\ell = 2$ is shown here). **(b)** The real-space field forms a cylindrically-symmetric vortex beam possessing the helical phase front and carrying the orbital AM $\langle \mathbf{L} \rangle \propto \ell \bar{\mathbf{z}}$. This angular momentum is produced by the spiraling canonical momentum density $\mathbf{p}$ in the beam (shown by cyan). Although all plane waves in the spectrum **(a)** are longitudinally polarized (i.e., the Fourier components of the velocity $\tilde{\mathbf{v}}(\mathbf{k}) \parallel \mathbf{k}$), the local polarization in real space, $\mathbf{v}(\mathbf{r})$, becomes elliptical, which produces a nonzero spin AM density $\mathbf{S} \propto \mathrm{Im}(\mathbf{v}^* \times \mathbf{v})$ (shown by red arrows).

To construct the acoustic Bessel beam, we will not calculate the Fourier integral involving the above plane-wave spectrum, but will use a simpler approach providing the same result [63]. We start with the well-known scalar part of the Bessel beam, described by the pressure field:

$$P = A J_\ell(\kappa r) \exp(i\ell\varphi + ik_z z). \tag{13}$$

Here, $A$ is a constant amplitude, $k_z = k\cos\theta_0$ is the longitudinal wave number, $\kappa = k\sin\theta_0$ is the transverse (radial) wave number, and $(r, \varphi, z)$ are the cylindrical coordinates in real space. Then, the vector velocity part of the acoustic Bessel field can be found from the second Eq. (3), $\mathbf{v} = -i(\rho\omega)^{-1} \nabla P$:

$$v_r = -iA' \frac{dJ_\ell(\kappa r)}{dr} \exp(i\ell\varphi + ik_z z) = -iA' \frac{\kappa}{2}\left[J_{\ell-1}(\kappa r) - J_{\ell+1}(\kappa r)\right]\exp(i\ell\varphi + ik_z z),$$

$$v_\varphi = A'\frac{\ell}{r}J_\ell(\kappa r)\exp(i\ell\varphi+ik_z z) = A'\frac{\kappa}{2}\big[J_{\ell-1}(\kappa r)+J_{\ell+1}(\kappa r)\big]\exp(i\ell\varphi+ik_z z),$$

$$v_z = A'k_z J_\ell(\kappa r)\exp(i\ell\varphi+ik_z z). \tag{14}$$

Here, $A' = A/(\rho\omega)$, and we used the cylindrical-coordinate components of the velocity, as well as the recurrence relations for the Bessel functions. The Cartesian components can be found as $v_x = v_r\cos\varphi - v_\varphi\sin\varphi$ and $v_y = v_r\sin\varphi + v_\varphi\cos\varphi$, which yields:

$$v_x = -iA'\frac{\kappa}{2}\big[J_{\ell-1}(\kappa r)e^{-i\varphi} - J_{\ell+1}(\kappa r)e^{i\varphi}\big]\exp(i\ell\varphi+ik_z z),$$

$$v_y = A'\frac{\kappa}{2}\big[J_{\ell-1}(\kappa r)e^{-i\varphi} + J_{\ell+1}(\kappa r)e^{i\varphi}\big]\exp(i\ell\varphi+ik_z z). \tag{15}$$

Equations (13)–(15) describe acoustic Bessel beams. Substituting these expressions into Eq. (5), and using and the dispersion relation (4), we find the energy density distribution:

$$W = \frac{\beta}{4}|A|^2\left\{(1+\cos^2\theta_0)J_\ell^2(\kappa r) + \frac{\sin^2\theta_0}{2}\big[J_{\ell-1}^2(\kappa r)+J_{\ell+1}^2(\kappa r)\big]\right\}. \tag{16}$$

This expression exhibits a phenomenon, which is also seen in electromagnetic [58,66] and Dirac-electron [59] vortex beams. In the paraxial limit, $\theta_0 \ll 1$, the radial distribution of the energy density is given by the single Bessel function as in the scalar case: $W(r) \propto J_\ell^2(\kappa r)$. In particular, it vanishes in the center of vortex beams: $W(0) \simeq 0$ for $\ell \neq 0$. However, for nonparaxial beams, when $\theta_0^2$-order terms are not negligible, the first-order vortex beams with $\ell = \pm 1$ have a *nonzero energy density in the center*: $W(0) = \beta|A|^2\sin^2\theta_0/8$, as can be seen in Figs. 2 and 3 below. To detect this nonzero energy density in the center of the first-order vortex beam, one needs a detector sensitive to the kinetic (velocity) part of the energy density. For example, the gradient acoustic force on a small particle is determined by the potential involving this kinetic energy density [67,68].

The energy flux density (5) or kinetic momentum density (7) acquire simple forms in acoustic Bessel beams (13)–(15):

$$\mathbf{\Pi} = c^2\frac{\beta}{2\omega}|A|^2\left(k_z\bar{\mathbf{z}} + \frac{\ell}{r}\bar{\boldsymbol{\varphi}}\right)J_\ell^2(\kappa r), \tag{17}$$

where the overbars indicate the unit vectors of the corresponding coordinates. The canonical momentum density (6) has a more sophisticated form. First, since all the Cartesian components of the Bessel-beam fields (13)–(15) share the same phase factor $\exp(ik_z z)$, it is easy to see that the $z$-component of the canonical momentum density (6) and the integral momentum of the Bessel beam can be written as:

$$p_z = k_z\frac{W}{\omega}, \qquad \frac{\omega\langle\mathbf{p}\rangle}{\langle W\rangle} = \frac{\omega\langle\mathbf{\Pi}\rangle}{c^2\langle W\rangle} = k_z\bar{\mathbf{z}}. \tag{18}$$

Here the purely longitudinal direction of the integral momentum follows from the cylindrical symmetry of the beam. Second, the canonical momentum density has the azimuthal $\varphi$-component, which determines the $z$-component of the orbital AM density (8). Substituting Eqs. (13)–(15) into Eqs. (6) and (8), we obtain these components:

$$L_z = p_\varphi r = \ell \frac{W}{\omega} - \frac{\beta |A|^2}{8\omega} \sin^2\theta_0 \left[ J_{\ell-1}^2(\kappa r) - J_{\ell+1}^2(\kappa r) \right]. \tag{19}$$

The spiraling streamlines of the momentum density $\mathbf{p} = p_z \bar{\mathbf{z}} + p_\varphi \bar{\boldsymbol{\varphi}}$, Eqs. (18) and (19), are schematically shown in Fig. 1(b).

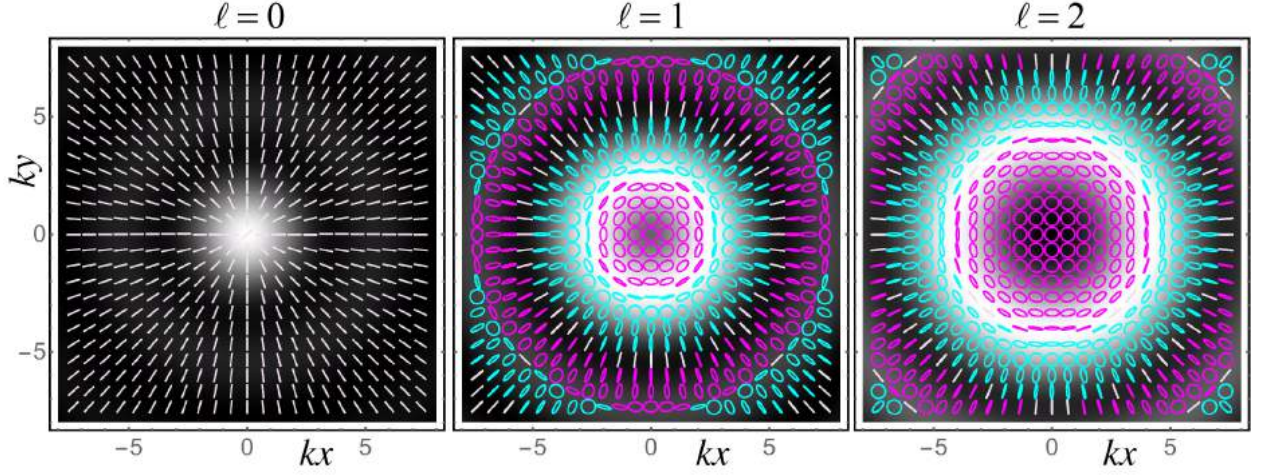

**Fig. 2.** Distributions of the energy-density $W(x,y)$ (greyscale background plots) and polarization ellipses of the velocity field $(v_x, v_y)$ in the transverse cross-sections of acoustic Bessel beams with $\theta_0 = \pi/4$ and different orders $\ell$. Magenta and cyan colors correspond to right-handed ($S_z > 0$) and left-handed ($S_z < 0$) elliptical polarizations, respectively. The non-vortex $\ell = 0$ beam is radially-polarized, i.e., $S_z \equiv 0$, while vortex beams with $\ell \neq 0$ have nonzero longitudinal spin density $S_z$ and purely circular polarization in the center. Flipping the sign of $\ell$ flips the handedness of the polarization, i.e., $S_z$. One can also see a nonzero energy density in the center of the $\ell = 1$ beam.

The spin AM density is also present in acoustic Bessel beams. Substituting the velocity components (14) and (15) into Eq. (7), we find the longitudinal and transverse (azimuthal) components of the spin AM density:

$$S_z = \frac{|A|^2}{\rho\omega^3} \frac{\ell}{r} J_\ell(\kappa r) \frac{dJ_\ell(\kappa r)}{dr} = \frac{\beta|A|^2}{4\omega} \sin^2\theta_0 \left[ J_{\ell-1}^2(\kappa r) - J_{\ell+1}^2(\kappa r) \right], \tag{20}$$

$$S_\varphi = -\frac{k_z |A|^2}{\rho\omega^3} J_\ell(\kappa r) \frac{dJ_\ell(\kappa r)}{dr} = -\frac{\beta|A|^2}{4\omega} \frac{k_z r}{\ell} \sin^2\theta_0 \left[ J_{\ell-1}^2(\kappa r) - J_{\ell+1}^2(\kappa r) \right]. \tag{21}$$

We first note that the longitudinal spin AM density (20) is induced by the vortex (proportional to $\ell$) and vanishes in the zero-order beam with $\ell = 0$. Similarly to electromagnetic waves, the $z$-component of the spin can be associated with the polarization ellipticity in the transverse $(x,y)$ plane. The transverse polarization distribution for the velocity field $(v_x, v_y)$ in acoustic Bessel beams is shown in Fig. 2. The $\ell = 0$ non-vortex beam has purely linear radial polarization, while vortex beams with $\ell \neq 0$ have radially-varying polarization, oscillating between the right-hand and left-hand circular polarizations. Notably, the polarization in the center of the vortex beam is always *purely circular*, with the handedness determined by $\text{sgn}(\ell)$. This can be seen from Eqs. (13) which yield $v_\varphi = i\,\text{sgn}(\ell)v_r$ for $r \to 0$. This phenomenon can be observed in the first-order beams with $\ell = \pm 1$, which have nonvanishing transverse velocity components in the center (while the longitudinal component $v_z$ vanishes). Remarkably, recent analysis of the acoustic torque on a small absorbing particle in a Bessel beam [65] show the radial dependence to be exactly proportional to $S_z$ in Eq. (20), including a non-vanishing torque in the center of the first-order Bessel beam. It is worth noticing that the circular transverse polarization in the center of the beam does not contradict the longitudinal character of the acoustic waves. The field (13) is curl-less, $\nabla \times \mathbf{v} = 0$, and its transverse polarization in the vortex center is a result of destructive interference of multiple purely-longitudinal plane waves with different wavevectors and phases, Fig. 1. This phenomenon is an acoustic counterpart of the abnormal longitudinal polarization in nonparaxial radially-polarized optical beams [69–72]. Next, the "transverse spin" (21) is quite similar to its optical counterpart, which appears in nonparaxial optical beams [5,6,11]. The sign of the transverse spin does not depend on the vortex sign, $\text{sgn}(\ell)$, and it does not vanish in the zero-order beam with $\ell = 0$. The distributions of the longitudinal-plane polarization $(v_z, v_r)$ in acoustic Bessel beams, with its ellipticity corresponding to the transverse spin (21), are shown in Fig. 3.

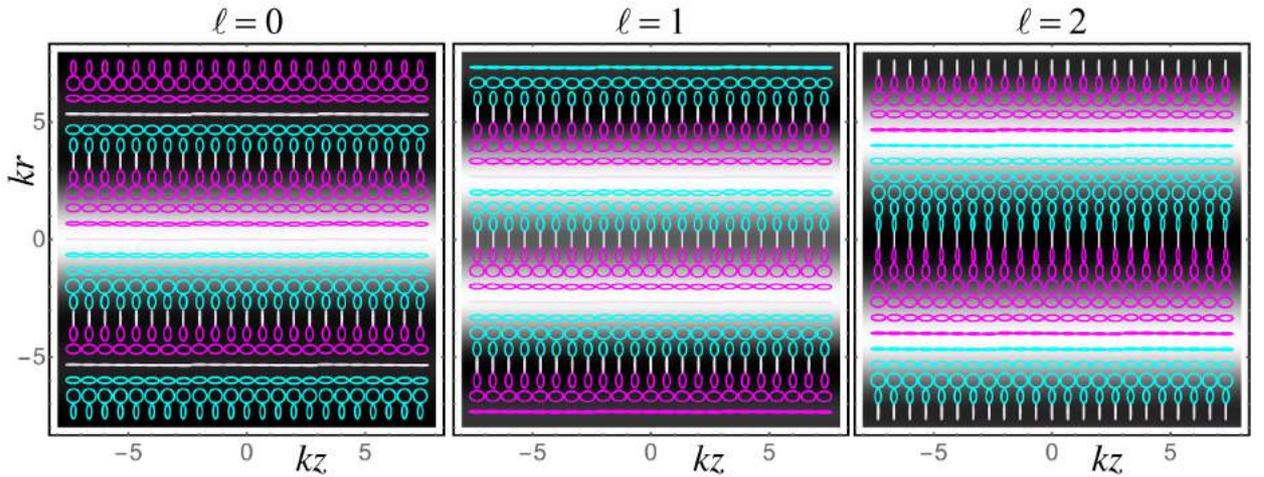

**Fig. 3.** Distributions of the energy-density $W(r,z)$ (greyscale background plots) and polarization ellipses of the velocity field $(v_r, v_z)$ in the longitudinal cross-sections of acoustic Bessel beams with $\theta_0 = \pi/4$ and different orders $\ell$. Magenta and cyan colors correspond to right-handed ($S_\varphi > 0$) and left-handed ($S_\varphi < 0$) elliptical polarizations, respectively. Flipping the sign of $\ell$ does not change the handedness of the polarization, i.e., does not affect the transverse spin $S_\varphi$.

Equations (19)–(21) reveal remarkable general properties of the spin and orbital AM in nonparaxial acoustic vortex beams. Namely, it follows from Eqs. (19) and (20) that

$$L_z + \frac{S_z}{2} = \left( \psi \left| \hat{L}_z + \hat{S}_z \right| \psi \right) = \ell \frac{W}{\omega}. \qquad (22)$$

This means that acoustic vortex beams are *eigenmodes of the "total AM" operator* $\hat{L}_z + \hat{S}_z$, but not eigenmodes of $\hat{L}_z$ and $\hat{S}_z$ separately. On the one hand, this is entirely similar to cylindrical optical [58,73,74] or Dirac-electron [59,61] modes, which have well defined integer *total* AM $\ell\hbar$ "per particle". This is a consequence of the intrinsic *spin-orbit interaction* in vector optical or quantum fields [58,59,75], and acoustic fields display similar features. On the other hand, there is a crucial difference due to the factor of 1/2 in the spin contribution in Eq. (21), which originates from the definition of the spin density as $\mathbf{S} = 2\langle\psi|\hat{\mathbf{S}}|\psi\rangle$ in Eq. (7). (Note that the normalized spin density $\omega|S_z|/W$ reaches the maximum value of 2 in the circularly-polarized center of the $|\ell|=1$ Bessel beams.) It is tempting to define the acoustic spin density as $\mathbf{S}' = \langle\psi|\hat{\mathbf{S}}|\psi\rangle$, which would yield $L_z + S_z' = \ell W/\omega$, but this contradicts the actual physical density of the intrinsic AM originating from the local elliptical motion of the medium particles [38–40]. Due to this, the *physical* total AM density (8) is not quantized and equals $J_z = L_z + S_z = \ell W/\omega + S_z/2$. Nonetheless, the general integral relations (9) hold true in acoustic Bessel beams, and the *integral* AM is properly quantized:

$$\frac{\omega\langle\mathbf{J}\rangle}{\langle W\rangle} = \frac{\omega\langle\mathbf{L}\rangle}{\langle W\rangle} = \ell\bar{\mathbf{z}}, \qquad \frac{\langle\mathbf{S}\rangle}{\langle W\rangle} = 0. \qquad (23)$$

Here, the purely longitudinal directions of the angular momenta follow from the cylindrical symmetry of the beam, while the vanishing of the spin contribution follows from Eq. (20) and relation $\ell \int_0^\infty J_\ell(\kappa r) \frac{dJ_\ell(\kappa r)}{dr} dr = 0$. Note that the non-integer character of the normalized total AM density, $\omega J_z/W \neq \ell$, and the integer character of the integral value (23) is in agreement with previous calculations [30,32] for nonparaxial acoustic vortex beams.

We finally note that the spin AM density (20) and (21) has its direction exactly *orthogonal* to the energy flux density (17) [which is a general fact following from Eqs. (5) and (7)], as schematically shown in Fig. 1(b):

$$\mathbf{S} \propto \left(\frac{\ell}{r}\bar{\mathbf{z}} - k_z\bar{\boldsymbol{\varphi}}\right), \qquad \mathbf{S}\cdot\boldsymbol{\Pi} = 0. \qquad (24)$$

The spin-density direction is described by the unit vector $\mathbf{s} = \mathbf{S}/|\mathbf{S}|$, which can be written as

$$\mathbf{s} = \sigma\frac{\ell\bar{\mathbf{z}} - k_z r\bar{\boldsymbol{\varphi}}}{\sqrt{\ell^2 + k_z^2 r^2}}. \qquad (25)$$

Here, the parameter $\sigma = \pm 1$ changes its sign with radius, such that $\sigma(r=0) = 1$ and it flips sign in every maximum or minimum of the radial distribution $J_\ell^2(\kappa r)$ for $r \neq 0$. Indeed, when $dJ_\ell^2(\kappa r)/dr = 0$, the spin density vanishes, $\mathbf{S} = 0$, see Eqs. (20) and (21), and its direction (25) is indeterminate (it is well-defined at $r=0$ because of the $1/r$ factor in the expression (20) for $S_z$). Note that the spin direction distribution in the vicinity of the beam axis resembles a *magnetic skyrmion* texture [76,77]. However, skyrmions have a smooth distribution of well-defined magnetization direction everywhere, while the distribution (25) has *singularities* (jumps to the opposite direction) at the $r \neq 0$ extrema of the radial distribution $J_\ell^2(\kappa r)$. These polarization singularities are the cylindrical "L-surfaces" of purely-linear polarization [78,79].

## 4. Discussion

We have considered properties of acoustic monochromatic fields in a homogeneous medium (fluid or gas). These are longitudinal waves described by one scalar (pressure) and one vector curl-free (velocity) fields. So far, mostly the energy and energy-flux densities were considered for these fields, and recently it was shown that inhomogeneous acoustic waves also possess a nonzero spin AM density [38–40]. We have described the whole set of dynamical characteristics of acoustic wave fields, analogues to their electromagnetic counterparts, including canonical momentum, spin momentum, orbital AM, and helicity densities. We have shown that the *helicity density vanishes identically* in acoustic fields, which reflects their non-chiral character. At the same time, the spin AM density is generically non-zero (the velocity field can have an elliptical local polarization), but the *integral* spin AM for localized acoustic fields *vanishes*. This is in agreement with the spin-0 nature of longitudinal phonons.

As an example of acoustic beams carrying both spin (locally) and orbital angular momenta, we have considered nonparaxial Bessel beams with the vortex of charge $\ell$. These beams exhibit a rather nontrivial polarization structure and a number of measurable spin-related phenomena, such as torque on small absorbing particles [39,65]. In particular, we have found that the energy density does *not* vanish in the center of the first-order, $\ell = \pm 1$, Bessel beams, which is similar to optical vortex beams and in contrast to scalar vortex beams. Moreover, the velocity polarization in the center of vortex beams is *purely circular*, with the handedness determined by the vortex sign, $\text{sgn}(\ell)$, and *transverse*, i.e., orthogonal to the beam axis. This is an acoustic counterpart of the *purely-longitudinal* polarization on the axis of optical radially-polarized focused beams [69–72]. Remarkably, the spin density in acoustic Bessel beams has both longitudinal and transverse (azimuthal) components, which are odd and even with respect to the vortex charge $\ell$, respectively. Also, the spin density vanishes and changes its direction to the opposite one in the $r \neq 0$ extrema of the radial Bessel-function distribution $J_\ell^2(\kappa r)$.

Importantly, all of the above phenomena are experimentally observable. The nonzero energy density in the center of the tightly-focused $\ell = \pm 1$ vortex beams can be measured by a detector sensitive to the kinetic (velocity) part of the energy density, cf. optical experiment [64]. The nonzero spin density can be directly measured via torques on small absorbing particles, similarly to the recent experiment [39] and calculations [65]. The main difficulty in such measurements is to discriminate between the torque from the spin AM and the torque from the orbital AM [32,80,81]. One can notice that the orbital torques always accompany the spin-related torques and these often act in the same direction. Therefore, an accurate theory of the acoustic field interaction with an absorbing particle and careful experiments are required to study acoustic spin- and orbital-induced torques.

The *T*-odd and *P*-even nature of the angular momentum and the local character of the spin density makes it the main natural property which can be coupled to an external *magnetic field* [5] (see, e.g., the Zeeman interaction, optical Faraday effect, magnetic circular dichroism, etc.). Therefore, it is natural to expect that the acoustic spin density will play an important role in *magneto-acoustic* effects. Note that magnetic circular dichroism [82], as well as interactions with the Zeeman-split atomic states [83,84], have been successfully employed to detect nontrivial local polarization and spin properties in inhomogeneous optical fields. Moreover, various magneto-acoustic interactions involving angular momentum of phonons in solids were recently intensively explored [25–27,85]. This marks magneto-acoustics as one of the most promising directions for future studies of acoustic spin-related phenomena.

We finally note that the separation of the spin and orbital degrees of freedom in electromagnetic waves is intimately related to the difference between the "canonical" and "kinetic" (Belinfante) pictures in relativistic field theory [8,42,45,86–89]. In this manner, the canonical momentum and angular momentum properties, as well as their local conservation laws, can be derived from the field Lagrangian and Noether's theorem. It would be highly desirable to develop similar field-theory approach to acoustic fields, to derive the quantities introduced in Eqs. (6)–(9), as well as their conservation laws, from the first principles.

**Acknowledgements:** We are grateful to A. Y. Bekshaev, J. Dressel, and M. A. Alonso for fruitful discussions and important corrections. This work was partially supported by MURI Center for Dynamic Magneto-Optics via the Air Force Office of Scientific Research (AFOSR) (FA9550-14-1-0040), Army Research Office (ARO) (Grant No. Grant No. W911NF-18-1-0358), Asian Office of Aerospace Research and Development (AOARD) (Grant No. FA2386-18-1-4045), Japan Science and Technology Agency (JST) (Q-LEAP program, ImPACT program, and CREST Grant No. JPMJCR1676), Japan Society for the Promotion of Science (JSPS) (JSPS-RFBR Grant No. 17-52-50023, and JSPS-FWO Grant No. VS.059.18N), RIKEN-AIST Challenge Research Fund, the John Templeton Foundation, and the Australian Research Council.